\documentclass[floatfix,aps,prd,eqsecnum,showpacs,amsmath,amssymb,nofootinbib,preprintnumbers,twocolumn]{revtex4}

\usepackage{graphicx}
\usepackage{color}
\usepackage{xspace}

\def\laq{~\raise 0.4ex\hbox{$<$}\kern -0.8em\lower 0.62ex\hbox{$\sim$}~}
\def\gaq{~\raise 0.4ex\hbox{$>$}\kern -0.7em\lower 0.62ex\hbox{$\sim$}~}

\def\beq{\begin{equation}}
\def\eeq{\end{equation}}
\def\bea{\begin{eqnarray}}
\def\eea{\end{eqnarray}}

\def \ra {\rightarrow}

\def \Mp {M_{\rm P}}

\def \Da {\Delta}
\def \da {\delta}
\def \b {\beta}
\def \a {\alpha}
\def \ap {\alpha^{\prime}}

\def \Ga {\Gamma}

\def \sg {\sigma}

\def \da {\delta}

\def \om {\omega}
\def \Om {\Omega}

\begin{document}

\preprint{BA-TH/706-16}

\title{Observable gravitational waves in pre-big bang cosmology: an update} 

\author{M. Gasperini$^{1,2}$}

\affiliation{
$^{1}$Dipartimento di Fisica, Universit\`{a} di Bari, Via G. Amendola
173, 70126 Bari, Italy\\
$^{2}$Istituto Nazionale di Fisica Nucleare, Sezione di Bari, Bari, Italy}

\begin{abstract}
In the light of the recent results concerning CMB observations and GW detection we address the question of whether it is possible, in a self-consistent inflationary framework, to simultaneously generate a spectrum of scalar metric perturbations in agreement with Planck data and a stochastic background of primordial gravitational radiation compatible with the design sensitivity of aLIGO/Virgo and/or eLISA. We suggest that this is possible in a string cosmology context, for a wide region of the parameter space of the so-called pre-big bang models. We also discuss the associated values of the tensor-to-scalar ratio relevant to the CMB polarization experiments. We conclude that future, cross-correlated results from CMB observations and GW detectors will be able to confirm or disprove pre-big bang models and -- in any case -- will impose new significant constraints on the basic string theory/cosmology parameters. 
\end{abstract}

\vspace {1cm}~


\maketitle

%%%%%%%%%%%%%%%%%%%%%%%%%%%%%%%%%%%%%%
%%%%%%%%%%%%%%%%%%%%%%%%%%%%%%%%%%%%%%

\section {Introduction}
\label{Sec1}
\setcounter{equation}{0}

%%%%%%%%%%%%%%%%%%%%%%%%%%%%%%%%%%%%%%
%%%%%%%%%%%%%%%%%%%%%%%%%%%%%%%%%%%%%%

It is known that a cosmological phase of growing space-time curvature and accelerated evolution (also called ``superinflation" \cite{1}), followed by a (non-singular) transition to the standard radiation-dominated regime, can produce a stochastic background of relic gravitational waves (GW) with a blue-tilted (i.e. growing with frequency) spectrum \cite{2,3}. In a string theory context such a type of cosmic evolution is naturally suggested by the underlying duality symmetries \cite{4}. The resulting string cosmology scenario -- the so-called pre-big bang scenario \cite{5} -- is typically characterized by a relic GW background with a large (and possibly detectable) intensity in the sensitivity band of present interferometric antennas \cite{6,7,8}(see \cite{9,10} for a comprehensive review, and \cite{11} for a recent discussion).

An obvious problem with this scenario is the associated amplification of scalar metric perturbations which -- like tensor perturbations -- are also characterized by a growing (and possibly steep) primordial spectrum. The pump-field controlling the evolution of the (canonically normalized) quantum fluctuations outside the horizon, and determining the primordial spectrum, is indeed the same for both scalar and tensor components of the metric perturbations \cite{12}. Such a growing spectral behavior is in clear conflict with all present observations concerning the  CMB radiation (see e.g. \cite{13}), whose anisotropy is naturally explained by the presence of scalar perturbations of the cosmic geometry with a {\em slightly decreasing} primordial spectrum (at least, up to the relevant pivot scale $k_\ast =0.05$ Mpc$^{-1}$).

A viable model, in the context of a growing-curvature inflationary scenario, thus requires that the metric perturbations directly amplified by inflation be efficiently suppressed at all (large) scales relevant to the CMB observations. In addition, the model has to be complemented with some additional mechanism able to produce an alternative background of adiabatic scalar perturbations with the required large-scale amplitude and the appropriate (slightly decreasing) spectrum. 

A simple and well-known known example of such a mechanism is provided by the amplification, dominance, and subsequent decay of the so-called ``curvaton" field \cite{14,15,16}. In the context of the pre-big bang scenario the role of the curvaton can be naturally played by the Kalb-Ramond axion $\sg$ \cite{17,18}, associated by space-time duality with the four-dimensional components of the NS-NS two-form appearing in the low-energy string effective action. In fact, the fluctuations of the axion field -- unlike the metric fluctuations -- can be amplified with a flat (or nearly flat) primordial spectrum \cite{19,20},  even in a background configuration typical of the pre-big bang phase (i.e. characterized by growing curvature, growing string coupling, and producing a growing spectrum of primordial metric perturbations). 

After the transition to the post-big bang regime the axion becomes massive and eventually decays, leaving an ``induced" distribution of adiabatic curvature perturbations whose spectrum has exactly the same behavior as the primordial axion spectrum. Depending on the model, and on the details of the (regular) ``bouncing" transition and of the (pre- and post-big bang) background evolution, these axion-induced scalar metric perturbations can thus meet all required properties and consistently explain the observed CMB anisotropy \cite{17,18}.

Given a model which satisfies all constraints needed to produce (via the axion/curvaton mechanism) a viable spectrum of scalar perturbations, a question which naturally arises is whether or not, in the same model, the associated amplification of tensor metric perturbations is efficient enough to produce a detectable GW background in the sensitivity band of existing gravitational antennas. The aim of this paper is to address this point (which has never been addressed in the past literature on this subject), and to provide an answer to the above question. 

Our main result is that, after imposing all relevant phenomenological constraints, there is still a wide allowed region of the parameter space where the intensity of the  stochastic GW background ($\Om_{GW}$), evaluated in the frequency bands accessible to Advanced LIGO/Virgo and/or to eLISA, is larger than the minimum intensity levels expected to be detectable by these antennas when operating at their final design sensitivity (see e.g. \cite{21,22}) . This means, respectively, $\Om_{GW} \geq 10^{-9}$ at a frequency $\om \sim 10^2$ Hz, and/or $\Om_{GW} \geq 10^{-13}$ at $\om \sim 10^{-2}$ Hz. This implies a concrete possibility, in the near future, of detecting the relic pre-big bang gravitons or -- in the absence of detection -- of imposing new significant constraints on the parameters of such a string cosmology scenario. 

An additional, related result of our discussion is the prediction of very low values for the parameter $r$, expressing the ratio of the tensor-to-scalar perturbation amplitudes, when $r$ is evaluated at the frequency scales relevant to the CMB polarization experiments. We find, more precisely,  $r \laq 0.01$ at the conventional pivot scale $k=0.05$ ${\rm Mpc}^{-1}$. 

The paper is organized as follows. In Sect. \ref{sec2} we introduce a class of inflationary models typical of the pre-big bang scenario, and discuss our main assumptions on the background parameters. In Sect. \ref{sec3} we impose on the models the  constraints needed for an efficient production of scalar metric perturbations compatible with present CMB observations. In Sect. \ref{sec4} we consider the associated spectrum of relic gravitational radiation and, after imposing all relevant conditions, we discuss the possible detection of GW of pre-big bang origin in the frequency bands of the interferometric antennas. In Sect. \ref{sec5} we compute the  tensor-to-scalar  ratio predicted by our class of models at the frequency scales relevant to the CMB polarization experiments. 
In Sect. \ref{sec6} we present our final remarks and conclusions.

%%%%%%%%%%%%%%%%%%%%%%%%%%%%%%%%%%%%%%
%%%%%%%%%%%%%%%%%%%%%%%%%%%%%%%%%%%%%%

\section{The model}
\label{sec2}
\setcounter{equation}{0}

The cosmological model we shall consider describes the accelerated, dilaton-driven evolution of a string background which, starting from the low-energy string perturbative vacuum, reaches the high curvature and strong coupling regime and eventually decays -- after a regular bounce -- into a four-dimensional state of standard, frozen-dilaton, radiation dominated evolution \cite{9,10}. 

Let us take into account the possible presence of extra dimensions, and their evolution during an initial phase of dynamical dimensional reduction. We shall work, in particular,  with a simple example of background geometry with three isotropically expanding dimensions and six (internal) shrinking dimensions, not necessarily isotropic, described in the so-called string frame by the metric
\beq
ds^2= dt^2 - a^2(t) |d\vec x|^2 - \sum_i b_i^2(t) dy_i^2, ~~ i=1, \dots, 6.
\label{21}
\eeq
This geometry is sourced by a dynamical dilaton field $\phi(t)$, and satisfies the cosmological equations following from the string effective action (see e.g. \cite{9,10}). The initial axion background is trivial, $\sg=0$, but its quantum fluctuations $\da \sg$ are nonvanishing.

We are interested, in this paper, in the amplification of the quantum fluctuations of the four-dimensional metric, $\da g_{\mu\nu}= h_{\mu\nu}$, and of the axion, $\da\sg$. Their evolution  is described by the canonical variables $u_h = \xi_h h$ and $u_\sg= \xi_\sg \da \sg$, whose Fourier components $u_k$ satisfy the standard mode equation \cite{12}
\beq u_k^{\prime \prime}- \left( k^2 - {\xi^{-1}\xi^{\prime\prime}} \right) u_k=0.
\label{22}
\eeq
Here a prime denotes differentiation with respect to the conformal time $\eta$ (such that $dt= a d\eta$), and $\xi_h$, $\xi_\sg$ are the so-called pump fields, which can be explicitly identified, for each type of perturbation, by writing the string effective action up to terms quadratic in the first-order fluctuations $h, \da \sg$, and imposing the canonical  (diagonalized) form on the kinetic part of the perturbed action. Working with the string frame configuration (\ref{21}) one obtains
\beq
\xi_h = a \left( \prod_{i=1}^6 b_i\right)^{1/2} e^{-\phi/2}
\label{23}
\eeq
for the metric perturbations \cite{9,10}, and 
\beq
\xi_\sg = a \left( \prod_{i=1}^6 b_i\right)^{-1/2} e^{\phi/2}
\label{24}
\eeq
for the perturbations of the Kalb-Ramond axion \cite{19,20}.
{In the absence of extra dimensions -- or in the case of extra dimensions frozen from the beginning -- the above pump fields reduce, respectively, to $\xi_h=a \exp{(-\phi/2)}$ and $\xi_\sg=a \exp{\phi/2}$.} 

Suppose now that in the phase of accelerated (inflationary) background evolution a given pump field, $\xi$, has a power-law behavior parametrized, in conformal time,  by $\xi (\eta) \sim (-\eta)^\a$, with $\eta<0$, $\eta \ra 0_-$. The primordial spectral distribution of the corresponding perturbation modes, $\da_k \equiv u_k/\xi$, leaving the horizon during that phase, and asymptotically normalized to a vacuum fluctuation spectrum, is then controlled by the power $\a$ as $\Delta^2(k) \equiv (k^3/2 \pi^2) |\da_k|^2 \sim k^{3-|2\a-1|}$  (see e.g. \cite{12}). Hence, to compute the spectra of interest for this paper, we need to specify the parametric evolution, in conformal time, of the background geometry and of the dilaton field during the different cosmological phases of the model we are considering. 

Let us separately discuss the pre-big bang and the post-big bang regimes.

%%%%%%%%%%%%%%%%%%%%%%%%%%%%%%%%%%%%%%
%%%%%%%%%%%%%%%%%%%%%%%%%%%%%%%%%%%%%%

\subsection{Pre-big bang parameters}
\label{sec2a}

We will consider a minimal model of pre-big bang (i.e. pre-bouncing) evolution, with two phases: an initial low energy and weak coupling phase, followed by a more ``stringy", high-curvature phase whose properties crucially depend on the inclusion of the so-called $\ap$ corrections into the string effective action \cite{23}.

In the initial phase, ranging in conformal time from $-\infty$ to a transition time scale $\eta=-\eta_s<0$, the background can be appropriately described by a vacuum solution of the tree-level string cosmology equations, with a Bianchi-I type metric \cite{24}
\beq
a(\eta) \sim (-\eta)^{\b_0\over 1- \b_0}, ~~~~ b_i(\eta) \sim  (-\eta)^{\b_i\over 1- \b_0}, ~~~ \eta<-\eta_s,
\label{25}
\eeq
and with 
\beq
\phi(\eta) \sim { \sum_i \b_i +3 \b_0 -1\over 1-\b_0} \ln (-\eta),
~~~ \eta<-\eta_s,
\label{26}
\eeq
where $\b_0$, $\b_i$ are constant parameters satisfying the Kasner-like condition
\beq
3 \b_0^2+ \sum_i \b_i^2=1.
\label{27}
\eeq

For this background the behavior of the metric pump field (\ref{23}) is always given by $\xi_h \sim (-\eta)^{1/2}$ (quite independently of the particular values of  $\b_0$, $\b_i$), and one obtains for the metric perturbations the (very steep) cubic primordial spectrum $\Da^2_h \sim k^3$ (modulo logarithmic corrections, see \cite{24,25}). For the axion pump field (\ref{24}) we have instead $\xi_\sg \sim (-\eta)^{\a_\sg}$, with $\a_\sg =(5 \b_0-1)/(2(1-\b_0)$, and we obtain the primordial spectral power \cite{9,10,11}
\beq
3-|2\a_\sg-1|= 3 -2\left|3 \b_0-1\over 1-\b_0\right|.
\label{28}
\eeq
Note that this spectral power  depends only on the parameter $\b_0$ controlling the evolution of the four-dimensional geometry (see Eq. (\ref{25})).

In the subsequent high-curvature string phase, ranging from $\eta=-\eta_s$ to the final strong-coupling scale $\eta=-\eta_1$ (with $\eta_s>\eta_1$), our background is described by a fixed-point solution of the string cosmology equations with higher-curvature $\ap$ corrections \cite{23}. For such solutions, which represent late-time attractors of the preceding low-energy evolution, the space-time curvature stays frozen at a constant scale $H_1$ controlled by the value of the fundamental string mass parameter, while the effective four-dimensional string coupling $g= (\prod_i  b_i)^{-1/2} \exp (\phi/2)$ has a growth which, in conformal time, can be described by a simple power-law behavior. Hence, for such solutions \cite{23},
\beq
a(\eta) \sim (-\eta)^{-1}, ~~~ g(\eta) \sim (-\eta)^{-\b},  ~~~
-\eta_s < \eta <-\eta_1,
\label{29}
\eeq
where $\b={\rm const} >0$.

From the definitions (\ref{23}), (\ref{24}) we immediately obtain the two pump fields, $\xi_h =a g^{-1} \sim (-\eta)^{-1+\b}$ and $\xi_\sg = a g \sim (-\eta)^{-1-\b}$, and the corresponding primordial spectral powers of  metric perturbations, $3-|3-2\b|$, and of  axion perturbations, $3-|3+2\b|=-2\b$, for the modes leaving the horizon during the string phase. {Once again we may note that these spectral powers are both determined by  a quantity -- the parameter $\b$ -- controlling the evolution in time of an effective four-dimensional variable, the effective string coupling $g(\eta)$.}

It is {also} important to note that the low-energy  definitions of the pump fields  (\ref{23}), (\ref{24}) can  be safely applied to the high-curvature  string phase, in spite of the presence of the $\ap$ corrections affecting the perturbation equation (\ref{22}), because we are considering a (fixed-point) background configuration characterized by constant values of the Hubble parameters and by linear evolution (in cosmic time) of the dilaton field. In such a case, in fact, the only change of the canonical perturbation equation with respect to its low-energy limit (\ref{22}) is an effective shift of the comoving frequency $k$, while the effective pump field -- and the associated spectral behavior -- keeps unchanged  (as discussed in \cite{26}).  

Summarizing we can say that, in the given model of pre-big bang evolution, the independent parameters of primary interest for this paper, i.e. the parameters which explicitly control the shape of the spectrum (amplitude, slope, and frequency position of the various branches), are four: $\b_0$, $\b$, $\eta_s$, $\eta_1$. The two parameters $\eta_s$ and $\eta_1$, determining the duration and the localization in time of the string phase, can be conveniently replaced by two equivalent (more physical) quantities: the Hubble parameter $H_1$, associated  with the curvature of the string phase, and the redshift parameter $z_s=a_1/a_s= \eta_s/\eta_1$, describing the expansion of the $3$-dimensional space during that phase. Finally, the parameter $\b$ can also be expressed in terms of the overall growth of the four-dimensional string coupling during the string phase, using the relation 
\beq
g_s/g_1=( \eta_s/\eta_1)^{-\b}=z_s^{-\b}.
\label{210}
\eeq

We may note that the model considered here is the same as the model already used to compare the pre-big bang production of gravitons and of primordial seeds for the cosmic magnetic fields \cite{27}, with  only one difference: in this paper, the value of $H_1$ will not be fixed a priori but will be left free to vary, consistently with the other parameters, to match the CMB  constraints on the scalar perturbation spectrum (see Sect. \ref{sec3}).

%%%%%%%%%%%%%%%%%%%%%%%%%%%%%%%%%%%%%%
%%%%%%%%%%%%%%%%%%%%%%%%%%%%%%%%%%%%%%

\subsection{Post-big bang parameters}
\label{sec2b}

For a full computation of the metric perturbation spectrum we need to specify the background model also in the post-big bang epoch of standard decelerated evolution and decreasing curvature, which follows the ``bouncing transition" expected to occur at the end of the string phase, when the background reaches the strong coupling regime. 

It should be recalled, in this respect, that the kinematic details of the bounce can possibly affect only the (very high) frequency modes crossing the horizon during the bouncing phase (as discussed in \cite{28,29} considering explicit, dilaton-based models of bouncing at scales smaller than Planckian, $H_1< \Mp= (8 \pi G)^{-1/2}$). For the applications of this paper we will neglect such a possible distortion of the spectrum (due to the bounce) near the end-point-frequency, assuming that the bounce is ``almost instantaneously" localized at the transition epoch $\eta=-\eta_1$. 

We shall also assume that in the post-big bang epoch the dilaton and the extra spatial dimensions are frozen {(see \cite{Deri} for a discussion of the post-bouncing behavior of the extra dimensions), that} the Universe is filled with a hot radiation gas, and -- most importantly -- that the axion background has the right properties and the dynamical behavior needed to implement the curvaton mechanism. Namely, that it is able to ``transform" the primordial spectrum of isocurvature axion fluctuations into a final spectrum of adiabatic and Gaussian scalar metric perturbations \cite{17,18}.

What we need, to this purpose, is that the axion background emerges from the bouncing transition with a mass $m$ and a non-trivial value $\sg_i \not=0$, displaced from the minimum of the (non-perturbative, periodic) axion potential. Excluding the exotic ``trans-Planckian" possibility $\sg_i>\Mp$ (possibly producing a phase of axion-dominated, post-big bang, slow-roll inflation \cite{18}), the effectve axion potential can be assumed (to a good approximation) to take the form of a quadratic mass term.

In such a case the time evolution of the axion becomes oscillating in time, with proper frequency $m$, as soon as the curvature drops below the scale $H_m \sim m$. The axion then behaves like a dust fluid, its energy density grows with respect to the radiation energy density, and eventually dominates the cosmic expansion at a curvature scale $H_\sg$ which depends on $\sg_i$, and which can be expressed as \cite{17,18} $H_\sg \sim m (\sg_i/\Mp)^4$. Finally, since the Kalb-Ramond axion is coupled to the electromagnetic field with gravitational strength, it tends to decay into photons at a rate $\Ga \sim m^3/\Mp^2$, and completely disappears from the cosmological scene at a curvature scale $H_d \sim \Ga$. To avoid disturbing the standard nucleosynthesis scenario with a dust-dominated phase, we expect the decay to occur prior to nucleosynthesis, namely at a scale $H_d$ such that $H_d >H_N \sim (1 {\rm MeV})^2/\Mp$.

For this type  of background evolution, the (minimal) coupling of the axion fluctuations $\da \sg$ to the metric perturbations automatically produces a super-horizon spectrum of adiabatic scalar perturbations, described by a Bardeen potential $\psi$ whose spectral modes $\psi_k$ are determined by the primordial axion modes as \cite{17,18} 
\beq
|\psi_k| \sim {\Mp\over \sg_i }|\da \sg_k|.
\label{211}
\eeq
However, given the decreasing of the space-time curvature during the post-big bang phase, and excluding the possibility that the axion is still subdominant at the decay epoch (i.e. that $H_d> H_\sg$, which would introduce ``non-Gaussian" properties in the final spectrum \cite{30}), we find that the above model can be consistently implemented only if the following hierarchy of curvature scales is satisfied: $H_1\geq H_m \geq H_\sg \geq H_d > H_N$, where $H_1<\Mp$. In Planck units:
\beq
{H_1\over \Mp} \geq {m\over \Mp} \geq {m\over \Mp}\left(\sg_i\over \Mp\right)^4 \geq \left(m\over \Mp\right)^3 >\left(1{\rm MeV}\over \Mp\right)^2 .
\label{212}
\eeq

The background evolution is of the matter-dominated type ($a \sim \eta^2$) in the axion phase ranging from $H_\sg$ to $H_d$, and of the radiation-dominated type ($ a\sim \eta$) in all other phases. Tensor perturbation modes re-entering the horizon during the axion phase will thus contribute to an ``axionic" branch of the relic GW spectrum with a slope different from those of the other branches, re-entering in the radiation-dominated phase. The effect of the axion phase, in particular, is to superimpose an additional ``red" tilt to the (typically ``blue") slope of the primordial tensor spectrum \cite{9,10}. Hence, the regime of post-big bang evolution  provide us with two more parameters to add to those determining the shape of the spectra of interest for this paper: the axion mass $m$ and the initial amplitude $\sg_i$.

Such parameters must satisfy the condition (\ref{212}) but, in practice, the particular values of $m$ and $\sg_i$ have a much smaller impact on our discussion (and on our final conclusions) than the pre-big bang parameters introduced in Sect. \ref{sec2a}. Firstly, they affect the amplitude but not the slope of the scalar perturbations produced via the curvaton mechanism: in fact, such a slope (whose value is of crucial importance to match the recent precision data on CMB anisotropy) is transferred unchanged to the scalar spectrum from the spectrum of the 
super-horizon axion fluctuations that are amplified by pre-big bang inflation,  and that are still outside the horizon at the epoch of axion dominance \cite{17,18}.

Secondly, as already mentioned, the values of $m$ and $\sg_i$ control the localization in time of the phase dominated by the oscillating axion, and thus  the 
 position in frequency of the associated change of slope in the tensor perturbation spectrum. Such a localization, however, does not seem to be of primary importance for a possible experimental detection of the relic GW background. The reasons are the following.

Let us notice, first of all, that the red tilt produced by the axion-dominated phase tends to create a local peak at the low-frequency end of the axion branch of the spectrum. Hence, choosing an early beginning of this phase ($\sg_i \ra \Mp$), i.e. moving the axion branch towards the high-frequency end of the tensor spectrum, on one hand tends to remove the peak away from the working frequency band of  the interferometric antennas ($10^{-2}$--$10^2$ Hz), and thus disfavor GW detection. On the other hand, this choice simultaneously depresses the amplitude of scalar metric perturbations (through Eq. (\ref{211})), and thus requires (in order to match CMB observations) higher values of the transition scale $H_1$ (see Sect. \ref{sec3}) which, in its turn, will enhance the overall amplitude of the GW spectrum (see Sect. \ref{sec4}).

On the contrary, choosing a ``late" occurrence of the axion-dominated phase ($\sg_i \ll \Mp$) has the opposite effects: the allowed values of $H_1$ are smaller, but the GW spectrum turns out to be locally enhanced at lower frequencies, possibly at (or near) the frequency window of the interferometers. In conclusion, the two effects tend to compensate each other, and the final result is not very sensitive to the given values of $\sg_i$ and $m$. 

For the illustrative purpose of this paper we will work with a simple configuration in which
\beq
\sg_i=\Mp, ~~~~~~ m=H_1,
\label{213}
\eeq
i.e. with a model in which the oscillating axion starts to become dominant directly at the beginning of the post-big bang era. In that case the background conditions (\ref{212}) are automatically satisfied provided that $1>H_1/\Mp> (1 {\rm MeV}/\Mp)^{2/3}$, namely for
\beq
1>{H_1\over \Mp} > 10^{-14}.
\label{214}
\eeq

Let us now discuss the constraints to be imposed on this model in order to obtain a viable spectrum of scalar metric perturbations.

%%%%%%%%%%%%%%%%%%%%%%%%%%%%%%%%%%%%%%
%%%%%%%%%%%%%%%%%%%%%%%%%%%%%%%%%%%%%%

\section{The relevant constraints from  CMB observations}
\label{sec3}
\setcounter{equation}{0}

The curvaton mechanism based on the Kalb-Ramond axion, and on the model introduced in the previous section, generates a super-horizon spectrum of adiabatic scalar metric perturbations which applies to all modes which are outside the horizon at the beginning of the axion-dominated phase \cite{17,18}. Namely,  to all modes with $\om<\om_\sg$, where $\om_\sg= H_\sg a_\sg/a$ is the proper frequency of the mode re-entering the horizon when $H=H_\sg$. (Throughout this paper we will work with the proper frequency $\om$, related to the Fourier parameter $k$ by $\om= k/a$. Also, we will use the notation $\a_\sg \equiv a(\eta_\sg)$, while $a \equiv a(\eta)$).

By taking into account that there are two phases of accelerated pre-big bang evolution, and by applying the results of previous computations \cite{17,18} (performed assuming the occurrence of a regular bouncing transition, strictly localized at the final epoch $\eta=-\eta_1$), we obtain a primordial spectrum of scalar perturbations with two different branches:
\bea
\Da_R^2(k)&=& {f^2(\sg_i)\over 2 \pi^2} \left(H_1\over \Mp\right)^2
\left(\om\over \om_1\right)^{-2\b}, ~~~ \om_s<\om<\om_\sg,
\nonumber \\
&=&
 {f^2(\sg_i)\over 2 \pi^2} \left(H_1\over \Mp\right)^2
 \left(\om_s\over \om_1\right)^{-2\b}\left(\om\over \om_s\right)^{3 -2\left|3 \b_0-1\over 1-\b_0\right|}, 
 \nonumber \\ &&
 \om<\om_s.
 \label{31}
 \eea
Here $\om_1 =H_1 a_1/a$ is the proper frequency of a mode crossing the horizon just at then end of the string phase, while $\om_s =H_s a_s/a$ is the proper frequency of a mode crossing the horizon at the beginning of the string phase (note that $H_s=H_1$, hence the ratio $\om_1/\om_s=a_1/a_s=z_s$ defines the reshift parameter related to the time extension of this phase). Modes with $\om>\om_s$ leave the horizon during the string phase, and the slope of their spectrum is determined by the power-law behavior of the string pump field (see Sect. \ref{sec2a}); modes with $\om<\om_s$ leave the horizon during the low-energy, dilaton-driven phase, and their slope is determined by the low-energy pump field. Finally, $f(\sg_i)$ is the transfer function connecting axion fluctuations to scalar metric perturbations: its explicit form has been numerically computed in \cite{18}, and in general is given by
\beq
f(\sg_i) \simeq 0.13 \,{\sg_i\over \Mp}+0.25 \,{\Mp\over \sg_i} -0.01.
\label{32}
\eeq
For the particular case $\sg_i=\Mp$ that we will consider here we thus obtain $f^2(\sg_i) \simeq 0.137$. Notice also that, in order to write the scalar spectrum in the form (\ref{31}), we have implicitly used the condition $\om_s <\om_\sg$, which is trivially satisfied for our choice of parameters (\ref{213}) (a choice which implies $H_\sg=H_1$ and $\om_\sg=\om_1$).

Let us now assume that the extension of the string phase towards the past is limited in time, in such a way that all (low-frequency) perturbation modes affecting the (large-scale) distances relevant to the observed CMB anisotropy leave the horizon during the preceding, low-energy, pre-big bang regime. More precisely, let us assume that
\beq
\om_\ast <\om_s,
\label{33}
\eeq
where $\om_\ast$ is the proper frequency corresponding to the pivot scale $k_\ast = 0.05$ Mpc$^{-1}$ to which CMB measurements are typically referred (see e.g. \cite{31}). In such a case the CMB observations directly constrain only the low-frequency branch of the spectrum (\ref{31}). (In the opposite case, $\om_s<\om_\ast$, we have explicitly checked that there is no significant overlap between the allowed region of parameter space satisfying the CMB constraints and the region compatible with a detectable GW background).

Assuming that Eq. (\ref{33}) is satisfied, and using the measured value of the scalar spectral index \cite{31,13}, $n_s\simeq 0.968$, we have thus to impose on the parameter $\b_0$ (according to Eq. (\ref{28})) the condition
\beq
3 -2\left|3 \b_0-1\over 1-\b_0\right| \equiv n_s-1 \simeq -0.032.
\label{34}
\eeq
It may be interesting to stress that this experimental constraints suggests (as recently pointed out also in \cite{11}) the existence of a small asymmetry between the rate of three-dimensional pre-big bang expansion and the rate of isotropic contraction of the six internal dimensions. A perfect scale-invariant spectrum, $n_s=1$, would be reproduced indeed by the Kasner-like solution (\ref{25})--(\ref{27}) with $\b_0=-\b_i=-1/3 =-0.333$. From the experimental constraint (\ref{34}) we obtain, instead, $\b_0 \simeq -0.348$, which implies, according to Eq. (\ref{27}), $\sum_i \b_i^2 \simeq 0.637$ (instead of $\sum_i \b_i^2 =2/3= 0.666$).

The second constraint to be imposed on the low-frequency branch of the spectrum (\ref{31}) comes from the normalization of the spectral amplitude at the pivot scale, and given by \cite{31,13} $\Da_R^2(k_\ast) \equiv A_s \simeq 3 \times 10^{-10}$. Using Eq. (\ref{34}) for the slope, the definition of $z_s$ for $\om_1/\om_s$, and the explicit computation of $\om_\ast/\om_s$ given in Appendix A, we can then write the normalization condition as a function of the two experimental inputs $n_s$, $A_s$, as follows:
\bea
&&
z_s^{2\b+n_s-1} \left(H_1\over \Mp\right)^{5-2n_s\over 2}
\left(m\over \Mp\right)^{1-n_s\over 3} \left(\sg_i\over \Mp \right)^{2(n_s-1)\over 3}
\nonumber \\ &&
= 2 \pi^2 f^{-2}(\sg_i) A_s 10^{27(n_s-1)}.
\label{35}
\eea
As anticipated in Sect. \ref{sec2}, we will use this condition to fix the transitions scale $H_1$ in terms of the other parameters $z_s$, $m$, $\sg_i$, in order to eliminate the unknown value of $H_1$ from all subsequent constraints.

Let us also recall that our assumption (\ref{33}) imposes a constraints on the duration of the string phase which (using Eq.(\ref{a4})) can be written as 
\beq z_s 
\left(H_1\over \Mp\right)^{-1/2}
\left(m\over \Mp\right)^{-1/3} \left(\sg_i\over \Mp \right)^{2/3}<10^{27}.
\label{36}
\eeq
In addition, the axion mass $m$ must satisfy the lower bound imposed by the nucleosynthesis (see Sect. \ref{sec2b}),
\beq
m > 10^{-14} \Mp.
\label{37}
\eeq
Finally, the string theory parameter $\b$ is not completely free, but it is constrained to be in the range
\beq 
0 \leq \b < 3.
\label{38}
\eeq
The lower limit on $\b$ is due to our assumption of growing string coupling (needed to implement a successful bouncing transition), while the upper limit is to be imposed to avoid background instabilities \cite{32}.

Summarizing, we can say that the considered model of background evolution may produce a viable spectrum of scalar metric perturbations, consistent with present CMB data, provide it satisfies the conditions (\ref{34})--({\ref{38}). In general, using the conditions (\ref{34}), (\ref{35}), we can always eliminate two parameters, and we are left with four independent quantities (for instance, the two string parameters $z_s$, $\b$ and the two axion parameters $m$, $\sg_i$). 

As discussed in Sect. \ref{sec2}, the two quantities $z_s$, $\b$ parametrize unknown (higher-derivative, strong coupling) string physics, and their particular values may have strong impact on the resulting phenomenology; the axion parameters, on the contrary, are less influential. Let us thus consider, as a simple illustrative example, the class of models where we fix $m$ and $\sg_i$, and we restrict our discussion to a string cosmology model with a two-dimensional parameter space spanned by $z_s$ and $\b$ -- or, equivalently, by $z_s$ and $g_s/g_1= z_s^{-\b}$.

We will use, in particular, the values of $m$ and $\sg_i$ suggested in Eq. (\ref{213}), putting everywhere $\sg_i=\Mp$ and $m=H_1$. Introducing the variables
\beq
x= \log z_s, ~~~~~~~ y= \log(g_s/g_1) \equiv - \b x,
\label{39}
\eeq
as convenient coordinates in our bi-dimensional parameter space, we can first obtain $H_1/\Mp$ in terms of $x$ and $y$ from the normalizations equation (\ref{35}), which gives
\beq
\log\left(H_1\over \Mp\right) = {6 \over 17- 5 n_s} \left[ 2y+ (1-n_s) x +27 n_s -34.4 \right],
\label{310}
\eeq
where $n_s=0.968$, and where we have used $A_s= 3 \times 10^{-10}$, and $\log [6 \pi^2 f^{-2}(\sg_i)] \simeq 2.6$. Using this constraints to eliminate $H_1/\Mp$ we can express the background condition (\ref{36}) as
\beq x-{5\over 6} y < 24,
\label{311}
\eeq
and the nuclesynthesis condition (\ref{37}) as
\beq
2y +(1-n_s)x > -5.2 -15.4 n_s.
\label{312}
\eeq
Finally, the allowed range (\ref{38}) of $\b$ implies the condition
\beq
0>y>-3x.
\label{313}
\eeq

The allowed region of the plane $(x,y)$ determined by the three inequalities (\ref{311})--(\ref{313}) will be compared, in the next section, with the allowed region compatible with the production of a stochastic GW background satisfying the existing phenomenological bounds, and possibly detectable by the interferometric antennas.

%%%%%%%%%%%%%%%%%%%%%%%%%%%%%%%%%%%%%%
%%%%%%%%%%%%%%%%%%%%%%%%%%%%%%%%%%%%%%

\section{The allowed parameter space of a detectable GW spectrum}
\label{sec4}
\setcounter{equation}{0}

Consider now the tensor perturbation modes which have been amplified in the context of the cosmological model illustrated in Sect. \ref{sec2}, and which are present today inside our horizon in the form of a stochastic background of relic gravitational waves. Their energy density evolves in time like the radiation energy density, and their frequency distribution is conveniently described (in units of critical energy) by the so-called spectral energy density, $\Om_{GW}(\om)$.

Such a spectrum has various branches, to be computed by solving the perturbation equation (\ref{22}) in the various phases of pre-big bang and post-big bang evolution, and by matching the corresponding solutions, assuming -- as stressed in Sect. \ref{sec2} -- the presence of a regular bouncing transition localized just at the end of the string phase, i.e. at $\eta=-\eta_1$. By recalling the results presented in  \cite{9,10}, and denoting with $\om_\sg$, $\om_d$ the proper frequencies of the modes re-entering the horizon, respectively, at the beginning and at the end of the axion-dominated phase, we find that the today value of the spectral energy density, $\Om_{GW}(\om, t_0)$, can be expressed in general as follows:
\bea
&& \!\!\!\!\!\!\!\!\!\!
\Om_{GW}(\om, t_0)= \Om_r\left(H_1\over \Mp\right)^2 
\left(\om\over \om_1\right)^{3-|3-2\b|},
\nonumber \\ &&
~~~ \om_\sg<\om<\om_1,
\nonumber \\ &&
=\Om_r \left(H_1\over \Mp\right)^2 
\left(\om_\sg\over \om_1\right)^{3-|3-2\b|}\left(\om\over \om_\sg\right)^{1-|3-2\b|},
\nonumber \\ &&
~~~ \om_d<\om<\om_\sg,
\nonumber \\ &&
=\Om_r \left(H_1\over \Mp\right)^2 
\left(\om_\sg\over \om_1\right)^{3-|3-2\b|}\left(\om_d\over \om_\sg\right)^{1-|3-2\b|}
\nonumber \\ &&
~~~\times \left(\om\over \om_d\right)^{3-|3-2\b|},
~~~~~~ \om_s<\om<\om_d,
\nonumber \\ &&
=\Om_r \left(H_1\over \Mp\right)^2 
\left(\om_\sg\over \om_1\right)^{3-|3-2\b|}\left(\om_d\over \om_\sg\right)^{1-|3-2\b|}
\nonumber \\ &&
~~~\times \left(\om_s\over \om_d\right)^{3-|3-2\b|} \left(\om\over \om_s\right)^3,
~~~~~ \om<\om_s.
\label{41}
\eea
Here $\Om_r$ is the present value of the total fraction of critical energy density in the form of cosmic radiation (and dominated by photons and neutrinos, see \cite{31} for its precise numerical value). 

As before, modes with $\om<\om_s$ leave the horizon during the low-energy, dilaton-driven phase, while modes with $\om >\om_s$ leave the horizon during the string phase. Their final spectral behavior, however, depends on if the primordial super-horizon modes re-enter the horizon during the radiation-dominated or the axion-dominated epoch. In the first case the slope of the primordial spectrum is unchanged, in the second case the slope is tilted towards the red by an additional $\om^{-2}$ factor \cite{9,10}.

As explicitly shown by the above spectrum, we have simplified the discussion of this paper by choosing a subclass of the general class of models introduced in Sec. \ref{sec3}, assuming that the post-big bang regime of axion-dominated evolution only affects the high-frequency modes leaving the horizon during the string phase. Namely, a subclass of models satisfying the background condition
\beq
\om_s< \om_d.
\label{42}
\eeq
Otherwise we should add to the above spectrum another possible (alternative) expression for $\Om_{GW}$, with $\om_d<\om_s$, where the frequency band modified by axion dominance concerns (in part, or totally) also the modes leaving the horizon during the low-energy dilaton-driven phase. We have not included this possible case since it would not enhance in a significant way the allowed region of parameter space compatible with GW detection.

We note, finally, that the spectrum (\ref{41}) is written in a form which applies to models with general values of $\sg_i$ and $m$. Our discussion, however, will be restricted to the particular case considered in Sect. \ref{sec3}, in which $\sg_i=\Mp$ and $m=H_1$. In that case $H_\sg=H_1$, $\om_\sg=\om_1$, and the frequency range of the first branch of the spectrum (\ref{41}) shrinks to a point, leaving the other three branches only. 

Let us now consider the phenomenological constraints to be imposed on the amplitude of the graviton spectrum. There are three main conditions.

A first condition comes from nucleosynthesis \cite{33}, not to spoil the accurate predictions on the abundance of light elements: the present value of the GW energy density, integrated over all modes and rescaled down to the nucleosynthesis epoch, cannot exceed, roughly, the energy density of one massless degree of freedom in thermal equilibrium (namely, about one tenth of the total energy density, see \cite{8} for a detailed computation). This bound can be translated into a crude upper limit on the peak intensity of the spectrum,
\beq
\Om_{GW}(\om_{\rm peak}) \laq 10^{-1} \Om_r,
\label{43}
\eeq
to be imposed at the peak frequency $\om_{\rm peak}$ (which, for the spectrum (\ref{41}), and for $\om_1=\om_\sg$, may correspond either to $\om_d$ or to $\om_\sg$, depending on the value of the parameter $\b$). 
Notice that this limit on $\Om_{GW}$ practically coincides with the upper bound recently placed by the LIGO and Virgo data \cite{34} on the amplitude of $\Om_{GW}$ in the frequency band of 41--169 Hz. 

A second, well known condition comes from the observations of millisecond pulsars \cite{35}: in particular, from the absence of any detectable distortion of pulsar timing due to the presence of a stochastic GW background, at a frequency scale $\om_p$ of the order of $10^{-8}$ Hz. This gives the bound
\beq
\Om_{GW}(\om_{p}) \laq 10^{-8} .
\label{44}
\eeq

A third important condition, for the class of models we are considering, comes from the request that the scalar metric perturbations directly amplified by the phase of pre-big bang inflation (with a primordial spectrum $\Da_\psi^2(k)$, closely related to the graviton spectrum (\ref{41})) be negligible with respect to the spectrum of adiabatic scalar perturbations $\Da_R^2(k)$ obtained from the axion (and given by Eq. (\ref{31})). The condition $\Da_\psi^2(k)< \Da_R^2(k)$ has to be satisfied in particular at the pivot scale $\om_\ast$, but, more generally, it has to be imposed at the maximum frequency scale $\om_M$ interested by the multipole expansion of the CMB anisotropy and constrained by present observations, namely at $\om =\om_M \simeq 6 \om_\ast$. Assuming that this scale still belongs  -- like $\om_\ast$ -- to the low-energy branch of the spectrum (\ref{41}), namely that  $\om_M<\om_s$, we thus obtain the condition:
\bea
&&
\left(H_1\over \Mp\right)^2 
\left(\om_\sg\over \om_1\right)^{3-|3-2\b|}\left(\om_d\over \om_\sg\right)^{1-|3-2\b|}
 \left(\om_s\over \om_d\right)^{3-|3-2\b|}
\nonumber \\ &&
~~~\times \left(\om_M\over \om_s\right)^3
< 2\pi^2 A_s  \left(\om_M\over \om_s\right)^{n_s-1}.
\label{45}
\eea

Thanks to this condition, the primordial scalar spectrum described by $\Da_\psi^2$ turns out to be negligible with respect to $\Da_R^2$ also at all lower frequency scales, since the decrease of $\Da_\psi^2$ with $\om$ is much faster than the decrease of $\Da_R^2$ (indeed, $\Da_\psi^2\sim \om^3$ while $\Da_R^2\sim \om^{n_s-1}$). We have checked, also, that this condition is slightly stronger than a similar constraint imposed on the ratio $r$ of the tensor-to-scalar spectral amplitude by the recent measurements of the CMB polarization \cite{36}, which imply, at the pivot scale, $r_{0.05} \equiv \Da_h^2(\om_\ast)/\Da_R^2(\om_\ast)<0.07$.

Summarizing, we can say that the considered model of pre-big bang inflation can consistently produce a relic stochastic GW background, compatible with known phenomenological bounds, provided the parameters of the spectrum (\ref{41}) satisfy the conditions imposed by Eqs. (\ref{42})--(\ref{45}). In addition, such a GW background is in principle accessible to the sensitivity of the aLIGO-AdVirgo detector network expected to be reached (in the 2020) by the so-called O5 observing run \cite{21}, provided that
\beq
\Om_{GW}(\om_L) \gaq 10^{-9} ,
\label{46}
\eeq
where $\om_L \simeq 10^2$ Hz.

The main object of this paper is to discuss the possibility that the same inflationary model producing a spectrum of scalar perturbations compatible with the CMB data -- and thus satisfying the previous constraints (\ref{310})--(\ref{313}) -- may consistently produce also a detectable GW background, satisfying all the constraints (\ref{42})--({\ref{46}). To this purpose we will determine the allowed region of parameter space
for the GW spectrum by imposing on the five independent parameters $\{z_s, \b, H_1, m, \sg_i\}$ not only the conditions (\ref{42})--({\ref{46}) presented in this section, but also the conditions (\ref{310})--(\ref{313}) derived in Sect. \ref{sec3} for a viable axion/curvaton scenario.

We will use the explicit form of the frequency ratios appearing in Eq. (\ref{41}) computed, for arbitrary values of the parameters, in Appendix A. We will then consider the same subclass of models as in Sect. \ref{sec3}, fixing the axion parameters to the values $\sg_i=\Mp$ and $m=H_1$, and restricting ourselves to a two-dimensional parameter space spanned by the variables $x$ and $y$ of Eq. (\ref{39}). Finally, we will eliminate everywhere $H_1$ through the condition (\ref{310}) (needed to match the experimental value of the scalar spectral amplitude). 
An example of the resulting graviton spectrum is illustrated in Fig. \ref{f1} for the particular case $x \equiv \log z_s =10$, and for three possible values of $\b\equiv-y/x$. Note that the ``end-point'' values  (i.e. the high-frequency cutoff and the associated amplitude) of the spectrum are controlled by $H_1$, and are thus strongly sensitive to the value of $y$ (according to Eq. (\ref{310})), as clearly shown by the figure. {Note also that, for particular values of the parameters, 
the obtained GW spectrum might mimic the one produced by a cosmological phase transition through the mechanism first discussed in \cite{phase,phase1} (see \cite{22} for a recent review of GW production from cosmological phase transitions).}

 %%%%%%%%%%%%%%%
\begin{figure}[t]
\includegraphics[width=8.6cm]{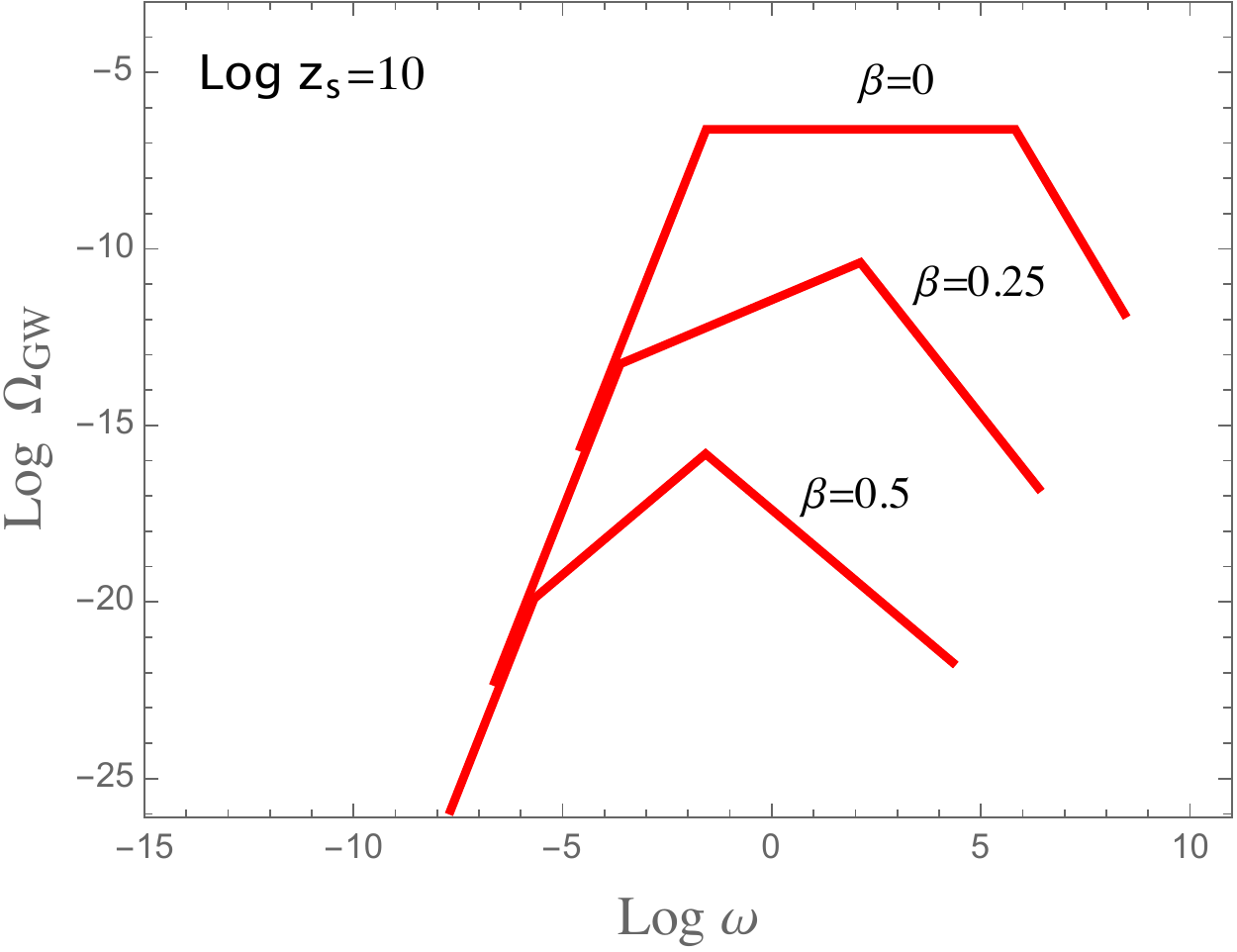}
\caption{A typical example of the GW spectrum (\ref{41}) for the considered class of models with  $\sg_i=\Mp$ and $m=H_1$, and with $H_1$  satisfying the CMB constraint (\ref{310}). We have used $\log \Om_r =-4$. We have considered the particular case $\log z_s =10$, and plotted the spectrum for three values of the string parameter, $\b =0$, $\b=0.25$ and $\b= 0.5$. The figure clearly illustrates the steep growth of the low-frequency branch of the spectrum ($\Om_{GW} \sim \om^{3}$ for $\om<\om_s$), the strong $\b$-dependence of the string branch of the spectrum ($\Om_{GW} \sim \om^{2\b}$ for $\om_s<\om<\om_d$), and the additional red tilt associated to the axion branch of the spectrum ($\Om_{GW} \sim \om^{2\b-2}$ for $\om_d<\om<\om_\sg$).} 
\label{f1}
\end{figure}
%%%%%%%%%%%%%%%%

Following the above procedure, and using Eq. (\ref{a6}) for the ratio $\om_d/\om_s$, we can then write the first constraint (\ref{42}) in the form
\beq
2y+{1\over 4}(21-9n_s)x>{1\over 4}(17-5n_s)+34.4-27n_s,
\label{47}
\eeq
and we obtain a first important result. 

It turns out, in fact, that this condition has  a vanishing intersection with the condition $\b>3/2$, represented in the $(x,y)$ plane by the inequality $y<-3x/2$. This means that, for the validity of the GW spectrum (\ref{41}), we must further restrict our class of models by replacing the range of $\b$ given in Eq. (\ref{38}) with the restricted range of values $0\leq \b < 3/2$, i.e. by replacing the condition (\ref{313}) with the more constraining condition
\beq
0>y>-{3x\over 2},
\label{48}
\eeq
to be used in all subsequent bounds on $\Om_{GW}$. 
This condition, by the way, helps simplifying our equations, since it implies $3-2\b>0$: hence, we can eliminate everywhere the modulus appearing in the powers of the spectrum (\ref{41}), putting $3-|3-2\b|=2\b$ and $1-|3-2\b|=2\b-2$.

Consider now Eq. (\ref{43}). As previously noted, the peak of our GW spectrum is located either at $\om_\sg$ or at $\om_d$, depending on whether the ``axion branch'' of the spectrum (i.e. the frequency band $\om_d <\om <\om_\sg$) is, respectively, growing ($1<\b<3/2$) or decreasing ($0<\b<1$) with frequency. In both cases, however, we have explicitly checked that the nucleosynthesis bound (\ref{43}) is always automatically satisfied, thanks to the more constraining background conditions (\ref{310})--(\ref{312}), (\ref{47}), (\ref{48}).

Then we move to the condition (\ref{45}), needed to depress the primordial scalar perturbations associated with the GW spectrum (\ref{41}). Using Eqs. (\ref{a6}), (\ref{a7}), and recalling that $\om_M \simeq 6 \om_\ast$, we find that such condition can be written as
\bea
&&
{5n_s-16\over 17- 5 n_s} \left[2y +(1-n_s)x +27 n_s - 34.4\right] 
\nonumber \\&& ~~~
+(4-n_s)x+2y 
< 97.4 -27n_s,
\label{49}
\eea
and that it imposes a further nontrivial limitation on the allowed region of  parameter space. 

We are left with the bound imposed by pulsar-timing data, Eq. (\ref{44}). In order to implement that bound we must separately consider the (alternative) cases in which the pulsar frequency $\om_p$ lies either in the low-energy branch, or in the string branch, or in the axion branch  of the GW spectrum (\ref{41}).

Using Eq. (\ref{a10}) we find that the first case  ($\om_p<\om_s$) is specified by  
\beq 
x <19.7 +{5\over 17-5 n_s} \left[2y +(1-n_s)x +27 n_s - 34.4\right] ,
\label{410}
\eeq
and that, in this case, the pulsar condition (\ref{44}) implies
\beq
-{11\over 17-5n_s}\left[2y +(1-n_s)x +27 n_s - 34.4\right] +3x+2y<55
\label{411}
\eeq
(we have used $\log \Om_r =-4$). 
The second alternative is defined by  $\om_p>\om_s$ and by  $\om_p<\om_d$ which, using Eq. (\ref{a12}), can be written as
\beq
{9\over 17-5n_s}\left[2y +(1-n_s)x +27 n_s - 34.4\right] >-19.7
\label{412}
\eeq
In this case the pulsar bound becomes
\bea
&&
{6\over 17-5n_s}\left({2\over 3}+{5y\over 3x}\right)\left[2y +(1-n_s)x +27 n_s - 34.4\right] 
\nonumber \\ && ~~~~~~
<-39.4{y\over x} -4.
\label{413}
\eea
Finally, the third case is identified by $\om_d<\om_p$, and the pulsar bound implies
\bea
&&
{6\over 17-5n_s}\left({11\over 3}+{5y\over 3x}\right)\left[2y +(1-n_s)x +27 n_s - 34.4\right] 
\nonumber \\ && ~~~~~~
<-39.4{y\over x} -43.4.
\label{414}
\eea

 %%%%%%%%%%%%%%%
\begin{figure}[t]
\includegraphics[width=8.6cm]{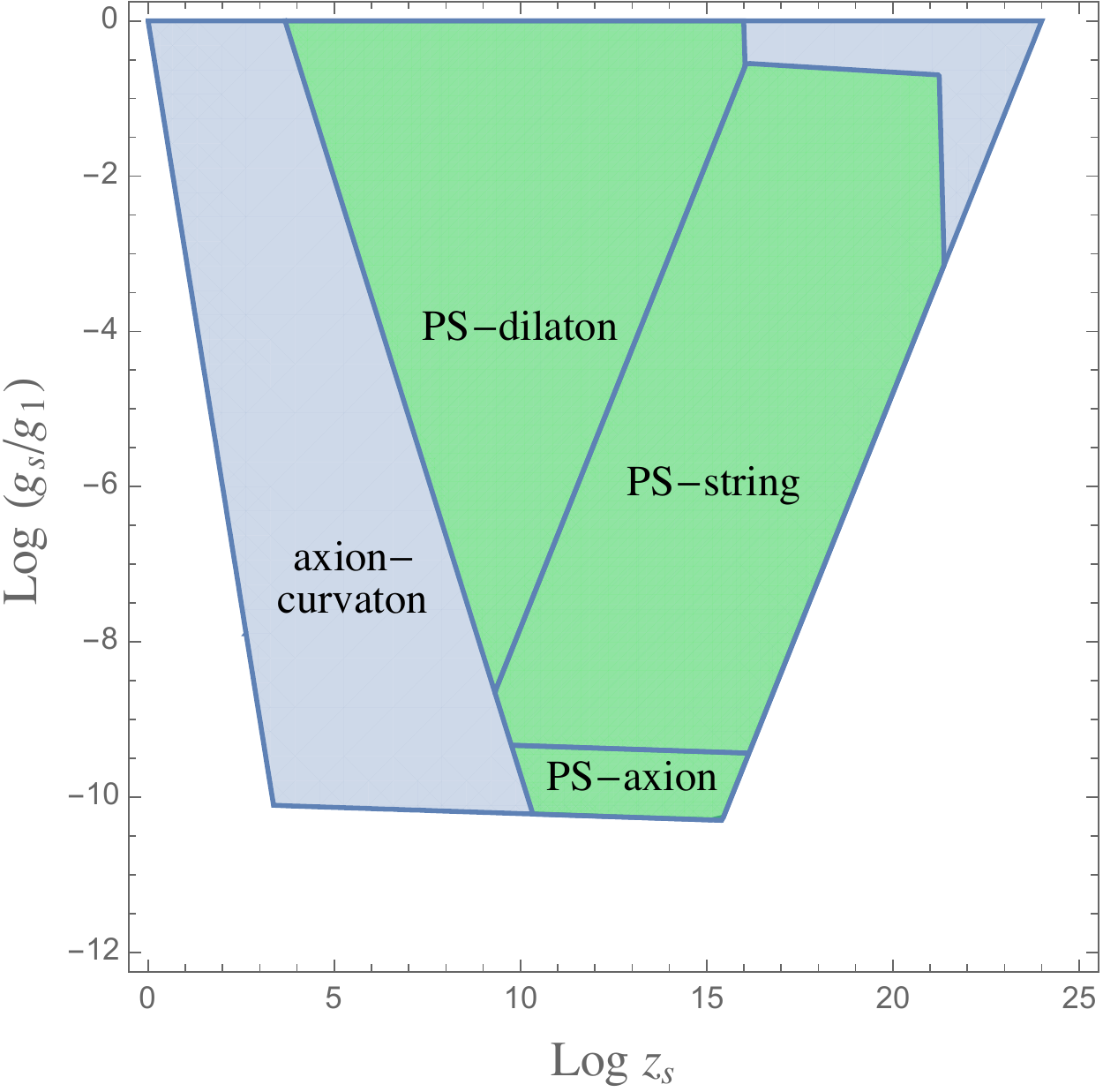}
\caption{The allowed values of the string cosmology parameters $z_s$ and $g_s/g_1$ for the production of a stochastic GW background which satisfies all phenomenological bounds (green area) and which is compatible with the production of a viable spectrum of scalar perturbations (light-blu trapezoidal area) via the axion/curvaton mechanism. The three green regions correspond to the three cases where the frequency $\om_p$ relevant to the pulsar constraint (\ref{44}) belongs either to the dilaton branch, or to the string branch, or to the axion branch of the spectrum (\ref{41}).} 
\label{f2}
\end{figure}
%%%%%%%%%%%%%%%%

The allowed region of the $(x,y)$ plane determined by Eqs. (\ref{47})--(\ref{414}) is shown (in green) in Fig. \ref{f2}, where it is superimposed to the (larger) allowed region  determined by Eqs. (\ref{310})--(\ref{313}) (the light-blue trapezoid). The green region defines the allowed values of the string cosmology parameters $\{z_s, g_s/g_1\}$ compatible with the production of a viable GW background and a viable spectrum of scalar metric perturbations. 

Let us now address the question of a possible detection of such a relic GW background, by adding to the previous constraints the condition (\ref{46}). Again we have to distinguish three cases, depending on the localization of the frequency 
$\om_L$ in the various branches of the spectrum (\ref{41}). 

If $\om_L<\om_s$, namely if (according to Eq. (\ref{a14}))
\beq 
x <9.7 +{5\over 17-5 n_s} \left[2y +(1-n_s)x +27 n_s - 34.4\right] ,
\label{415}
\eeq
then the maximum sensitivity of the aLIGO-AdVirgo network is attained in the dilaton branch of the spectrum, and the GW signal is detectable, according to Eq. (\ref{46}),  provided that 
\beq
-{11\over 17-5n_s}\left[2y +(1-n_s)x +27 n_s - 34.4\right] +3x+2y>24.
\label{416}
\eeq
If $\om_L>\om_s$ and, in addition  $\om_L<\om_d$ which means,  according to Eq. (\ref{a15}),
\beq
{9\over 17-5n_s}\left[2y +(1-n_s)x +27 n_s - 34.4\right] >-9.7,
\label{417}
\eeq
then the maximum sensitivity is in the string branch of the spectrum, and the GW background is detectable provided that
\bea
&&
{6\over 17-5n_s}\left({2\over 3}+{5y\over 3x}\right)\left[2y +(1-n_s)x +27 n_s - 34.4\right] 
\nonumber \\ && ~~~~~~
> -19.4{y\over x} -5.
\label{418}
\eea
Finally, if $\om_L>\om_d$, then the maximum expected sensitivity of the Earth-based interferometers corresponds to the axion branch of the spectrum (\ref{41}). In that case an explicit check shows that the detection condition (\ref{46}) cannot be satisfied, being incompatible with the set of all previous constraints (quite independently of the localization of the pulsar frequency $\om_p$ in the various branches  of the graviton spectrum). Hence, in that case, detection is impossible, unless the future advanced interferometers will reach values of sensitivity higher than those currently expected, thus relaxing the limiting condition (\ref{46}) (as will be discussed later at the end of this section).

% %%%%%%%%%%%%%%%
\begin{figure}[t]
\includegraphics[width=8.6cm]{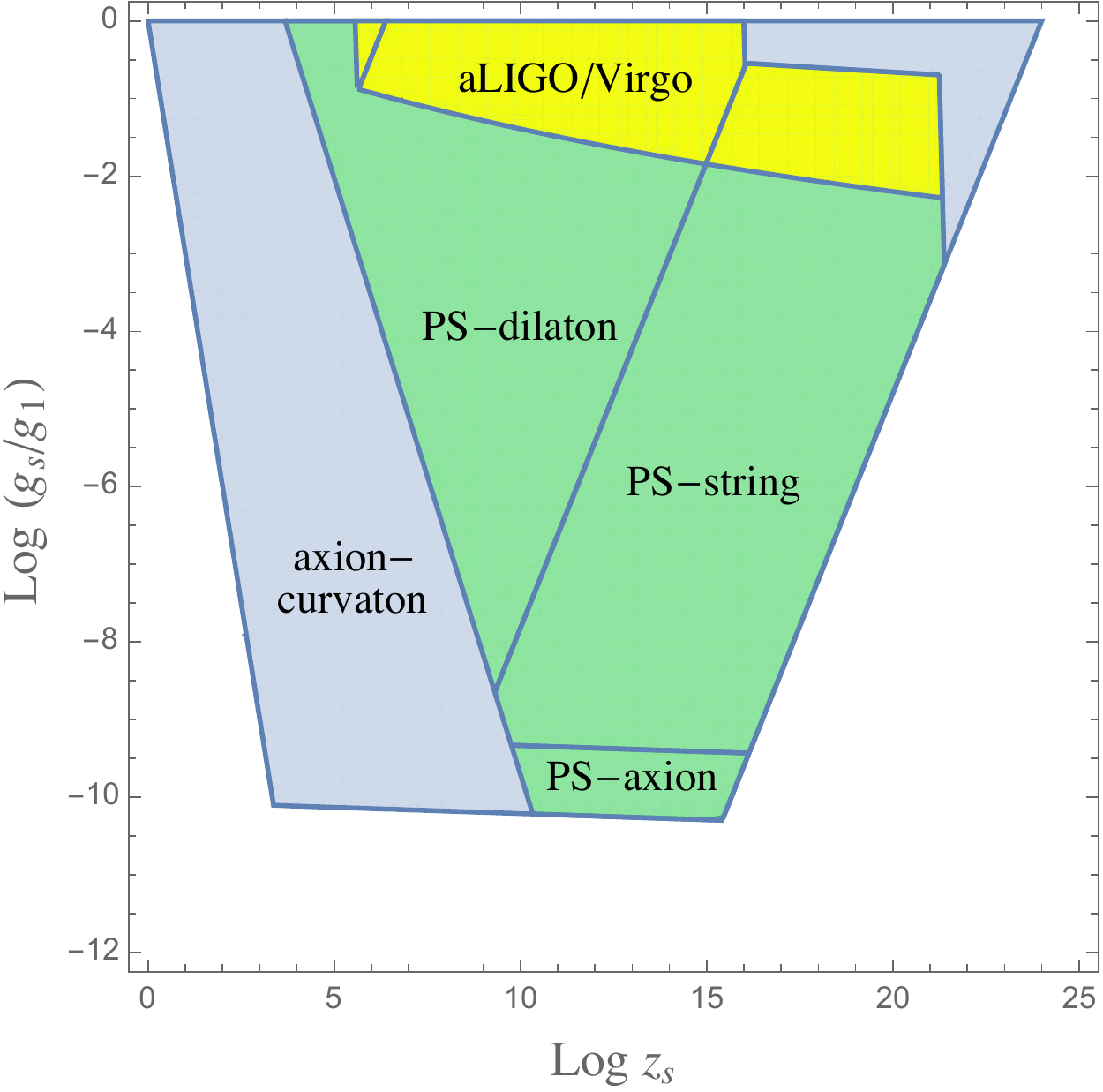}
\caption{The yellow area defines the allowed values of the string cosmology parameters $z_s$ and $g_s/g_1$ compatible with the production of a viable GW background which satisfies all required constraints (the green area of Fig. \ref{f1}) and which, in addition, may be detectable by the aLIGO/Virgo network. The three yellow regions corresponds to the detection condition (\ref{46}) imposed on the dilaton branch (the smallest region, on the left) and on the string branch (the two larger regions, on the middle and on the right) of the graviton spectrum (\ref{41}).} 
\label{f3}
\end{figure}
%%%%%%%%%%%%%%%%

Detection is possible, instead, at the sensitivity level of Eq. (\ref{46}), for the two cases described by Eqs. (\ref{415})--(\ref{418}), which are compatible with all the phenomenological constraints to be imposed on the spectrum. We may note, in particular, that if $\om_L$ is in the dilaton branch of the spectrum then also the pulsar frequency has to be in the same low-energy branch (since $\om_p<\om_L$); but if $\om_L$ is the string branch, then the detection is compatible with the pulsar bound (\ref{44}) imposed either in the dilaton or in the string branch of the graviton spectrum.

The results of the above discussion are graphically illustrated in Fig. \ref{f3}, where we have superimposed to the allowed region of a viable graviton spectrum (the green area of Fig. \ref{f1}) the region defined by the additional constraints (\ref{415})--(\ref{418}), needed for its detection. The resulting area (in yellow) describes the region of parameter space characterizing a class of string cosmology models able to produce a viable spectrum of scalar metric perturbations and a background of relic GW which is viable {\em and detectable} at the sensitivity level of Eq. (\ref{46}). 

%%%%

% %%%%%%%%%%%%%%%
\begin{figure}[t]
\includegraphics[width=8.6cm]{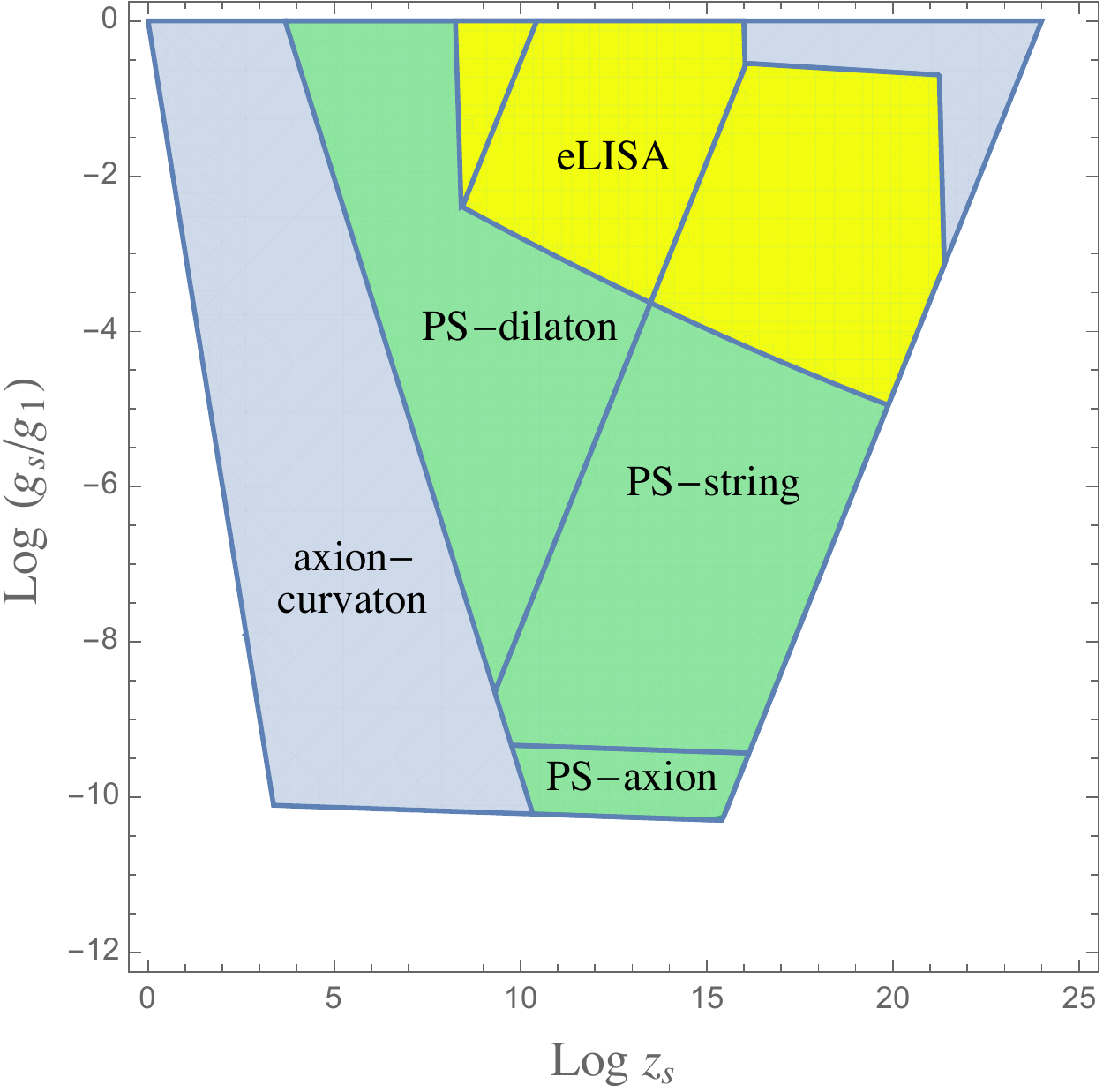}
\caption{Same as Fig. \ref{f3} but with the LIGO/Virgo condition (\ref{46}) replaced by the condition (\ref{419}), which has to be satisfied by a GW background expected to be detectable by eLISA in the so-called C1 configuration \cite{22}.} 
\label{f4}
\end{figure}
%%%%%%%%%%%%%%%

The above discussion can be easily repeated for the case of space-based interferometers such a eLISA, taking into account that the maximum sensitivity is attained in the frequency band $\om_{eL} \simeq 10^{-2}$ Hz. In particular, a stochastic GW background is expected to be detectable by eLISA in the so-called C1 configuration \cite{22} provided that 
\beq
\Om_{GW}(\om_{eL}) \gaq 10^{-13} .
\label{419}
\eeq

By replacing the LIGO/Virgo condition (\ref{46}) with the above condition, and applying the same procedure as before, we find again that the detection is incompatible with the other bounds if $\om_{eL}$ belongs to the axion branch of the spectrum ($\om_{eL}>\om_d$). In the other cases  ($\om_{eL}<\om_d$)  the detection is allowed, in the sense that the condition (\ref{419}) can be consistently satisfied together with all the phenomenological bounds needed to obtain a viable GW spectrum and a viable spectrum of scalar perturbations.

%%%%%%%%%%

The allowed region of parameter space determined by Eq. (\ref{419}) is illustrated in Fig. \ref{f4} (the yellow area), where it is also  compared with the regions allowed by the other phenomenological bounds. We can notice that there is a large overlap between the regions compatible with detection by aLIGO/Virgo (Fig. \ref{f3}) and by eLISA (Fig. \ref{f4}), but there are also important physical differences. 

In fact, the space-based detectors are sensitive to a smaller range of variation of the string parameter $z_s$ than the Earth-based detectors; however, they are compatible with a much larger range of variation of the  string-coupling ratio $g_s/g_1$. 

This implies that, to be detectable by eLISA at the sensitivity level of Eq. (\ref{419}), the GW background must be produced by a regime of pre-big bang evolution with a ``long enough" string phase at high curvature ($\log z_s >8$). Conversely, to be detectable by aLIGO/Virgo at the sensitivity level of Eq. (\ref{46}), the GW background must be produced by a pre-big bang regime with a ``small enough" growth of the string coupling during the high-curvature phase ($\log (g_s/g_1) \gaq -2$).

Let us discuss, finally, how the allowed region of parameter space compatible with GW detection is expected to vary with the variation of the level of experimental sensitivity to a stochastic GW background.  

Let us consider, to this purpose, the frequency band $\om_L \simeq 10^2$ Hz of the Earth-based interferometers, and let us repeat exactly the same procedure reported before to discuss the detection of the produced GW by the aLIGO/Virgo network, imposing however the detection condition (\ref{46}) with different values of the limiting sensitivity. We will use, for our illustrative purpose, the three different conditions $\Om_{GW} (\om_L) >10^{-7}$, $\Om_{GW} (\om_L) >10^{-9}$, and $\Om_{GW} (\om_L) >10^{-11}$. 

The second condition exactly reproduces our previous results, described by the yellow allowed region of Fig. \ref{f3}. The first condition is more constraining, and is incompatible with the pulsar bound (\ref{44}) imposed on the string branch of the GW spectrum.  The corresponding allowed region is smaller, and included into the yellow region of Fig. \ref{f3}. The third condition is less constraining, and is compatible with GW detection even if the sensitivity band $\om_L$  corresponds to the axion branch of the spectrum (for the pulsar frequency $\om_p$ localized either in the dilaton or in the string branch of the spectrum). The associated allowed region is larger, and obviously includes the yellow region of Fig. \ref{f3}. 

The above results are quantitatively illustrated in Fig. \ref{f5}, where we have plotted in red the allowed region determined by the condition $\Om_{GW} >10^{-11}$, we have plotted (and superimposed) in yellow the allowed region determined by the condition $\Om_{GW} >10^{-9}$, and we have plotted (and superimposed) in orange the allowed region determined by the condition $\Om_{GW} >10^{-7}$.  The considered example precisely shows how the allowed region of parameter space shrinks or expands with the variation of the level of the experimental sensitivity. A similar exercise can be easily performed also in the frequency band of eLISA, $\om_{eL}=10^{-2}$ Hz. 

% %%%%%%%%%%%%%%%
\begin{figure}[t]
\includegraphics[width=8.6cm]{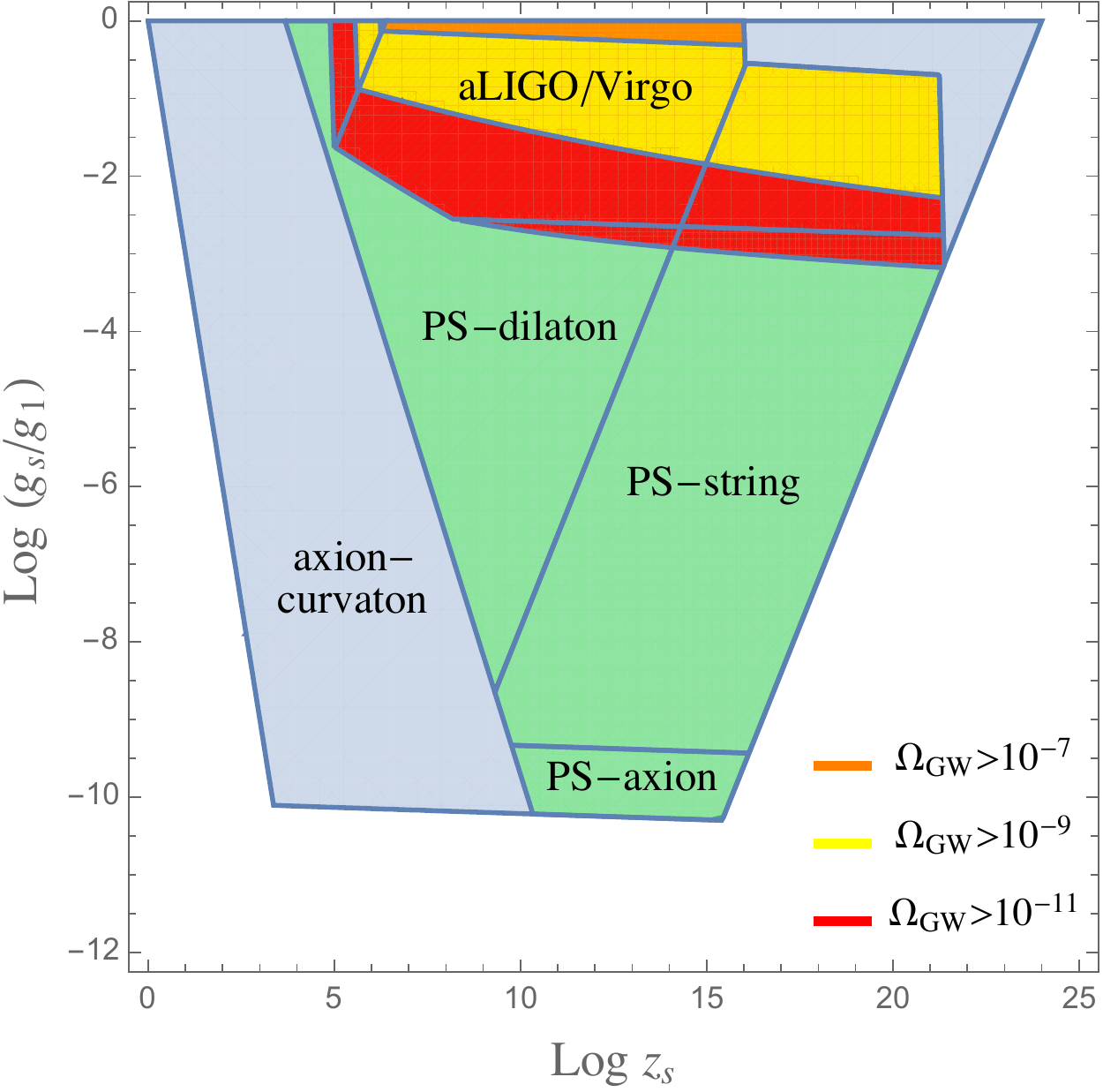}
\caption{Same as Fig. \ref{f3}, but for three different values of the limiting experimental sensitivity in the band of LIGO/Virgo. In particular, for $\Om_{GW} >10^{-7}$ (orange area), $\Om_{GW} >10^{-9}$ (yellow area), and $\Om_{GW} >10^{-11}$ (red area).} 
\label{f5}
\end{figure}
%%%%%%%%%%%%%%%%

%%%%%%%%%%%%%%%%%%%%%%%%%%%%%%%%%%%%%%
%%%%%%%%%%%%%%%%%%%%%%%%%%%%%%%%%%%%%%

\section{The predicted values of the tensor-to-scalar ratio}
\label{sec5}
\setcounter{equation}{0}

Let us conclude our discussion with an important remark concerning the predicted value of a measurable quantity, the tensor-to-scalar ratio $r$ evaluated at the pivot scale $\om= \om_\ast$.

Considering the primordial tensor spectrum corresponding to Eq. (\ref{41}), and using the results of Appendix A, we immediately obtain
\beq
r_{0.05} \equiv r(\om_\ast) ={\Da^2_h(\om_\ast)\over A_s}={10^{-81}\over 2 \pi^2 A_s}\left(H_1\over \Mp\right)^{-11/6} z_s^{3-2\b},
\label{420}
\eeq
where $A_s \simeq 3 \times 10^{-10}$ is the experimental value of the scalar spectral amplitude at $\om=\om_\ast$ \cite{31}. By eliminating $H_1$ through Eq. (\ref{310}) (to be consistent with the production of a realistic spectrum of scalar perturbations), we then find that the allowed values of $r_{0.05}$ predicted by our model can be expressed as
\bea
\log r_{0.05} &=&-{11\over 17-5n_s} \left[2y +(1-n)x +27n_s -34.4\right] 
\nonumber \\ &+&
3x+2y-72.8. 
\label{421}
\eea

% %%%%%%%%%%%%%%%
\begin{figure}[t]
\includegraphics[width=8.6cm]{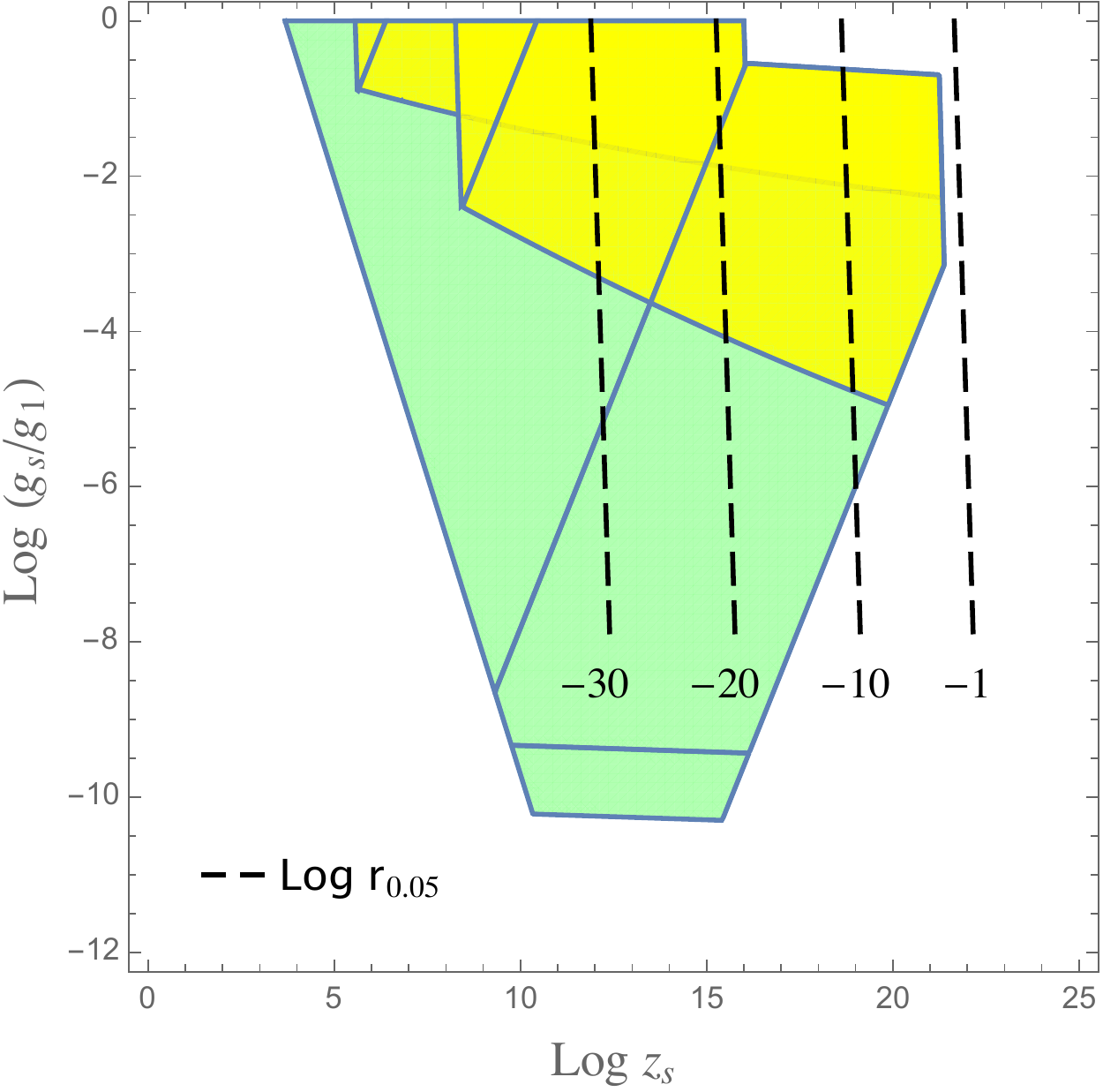}
\caption{The dashed lines describes the position in parameter space of four possible values of $r_{0.05}$ predicted by our class of models, consistently with the production of a viable spectrum of scalar metric perturbations. The lines are obtained from Eq.  (\ref{421}) with $\log r_{0.05} =-1, -10, -20, -30$. The green area is the same as in the previous figures. The yellow area describes the union of the two yellow regions of Figs. \ref{f3} and \ref{f4}. The right border of the yellow area corresponds to the dashed line with $\log r_{0.05} \simeq -2$.} 
\label{f6}
\end{figure}
%%%%%%%%%%%%%%%

Using this equation, we have plotted various curves at different, constant values of $r_{0.05}$ in the region of the $(x,y)$ plane compatible with the production of a viable -- and possibly detectable -- GW background. The results are illustrated in Fig. \ref{f6}, where the green area is the same as the one of the previous figures, while the yellow area describes the union of the two yellow regions of Fig. \ref{f3} and \ref{f4} (namely, it corresponds to the full region of parameter space compatible with the direct detection of the GW background either by aLIGO/Virgo or by eLISA at their design sensitivity, and determined, respectively, by the conditions (\ref{46}) and (\ref{419})).

As clearly shown by the dashed lines of Fig. \ref{f6}, only very small values of $r_{0.05}$ are compatible with the model of pre-big bang inflation considered in this paper. In particular, our class of models predicts
\beq
r_{0.05} \laq 0.01
\label{422}
\eeq
since, to fall inside the {green} allowed region, the dashed lines must satisfy the condition $\log r_{0.05} \leq -2$. It may  be appropriate to recall here that the most recent  experimental data \cite{36} imply $r_{0.05} \leq 0.07$, i.e. $\log r_{0.05} \laq -1.15$.

It should be noted, finally, that a small but possibly detectable (by future experiments) value of $r_{0.05}$, satisfying the limit (\ref{422}), is not incompatible in principle with a direct detection of the GW background by the interferometric antennas, provided the regime of pre-big bang inflation is characterized by a long enough high-curvature string phase (namely,  by values of $z_s$ corresponding to regions which are near enough to the right border of the yellow region of Fig. \ref{f6}).

%%%%%%%%%%%%%%%%%%%%%%%%%%%%%%%%%%%%%%
%%%%%%%%%%%%%%%%%%%%%%%%%%%%%%%%%%%%%%

\section{Conclusion}
\label{sec6}
\setcounter{equation}{0}

The standard (slow-roll) inflationary scenario naturally generates scalar perturbations which beautifully explain present data on CMB anisotropies; however, it predicts in general a very low, undetectable background of relic GW in the sensitivity band of present interferometric antennas\footnote{Only future space-based interferometers, like those planned in the context of the DECIGO \cite{37} and/or BBO \cite{38} projects, are in principle able to approach the required sensitivity levels.}. On the contrary, the pre-big bang inflationary scenario can explain CMB observations through the axion/curvaton mechanism and simultaneously produce a GW background strong enough to be observable by the aLIGO/Virgo and eLISA networks. 

Hence, future cross-correlated results from CMB experiments and GW detectors will provide us with a unique possibility of testing the cosmological dynamics for a better localization in time of the inflationary phase, and a more precise reconstruction of the past  history of our cosmo. 

In particular, {if we were to measure (through the B-mode polarization experiments) a value of $r_{0.05}$ larger (or not much smaller) than the limit (\ref{422}), without directly detecting a GW signal in the frequency range of present antennas at their final design sensitivity,} we would obtain a strong indication that the tensor perturbation spectrum is (on the average) decreasing over a wide range of frequencies, and that the primordial gravitons have been produced by a phase of standard inflation localized {\em after} the epoch of the big bang (see however \cite{Ben} for a possibility to evade this conclusion). 

Conversely, if we were to directly detect the GW background of inflationary origin in the frequency bands of the interferometers,
{with no signal, or a very low signal (satisfying the bound (\ref{422})) of tensor perturbations in the CMB polarization at large scales,} we would obtain an indication that the tensor spectrum on the average is growing with frequency, and a suggestion that the observed gravitons might have been produced {\em before} the bouncing transition which may play the role of the big bang in a  string cosmology context. 

Let us thus wait  for the answers which, as usual in physics, will be provided by the more and more precise forthcoming experiments. 

%%%%%%%%%%%%%%%%%%%%%%%%%%%%%%%%%%%%%%
%%%%%%%%%%%%%%%%%%%%%%%%%%%%%%%%%%%%%%

\section*{ACKNOWLEDGMENTS}

It is a pleasure to thank Gabriele Veneziano for his comments and suggestions that led to improve a preliminary version of this paper. I am also grateful to the Galileo Galilei Institute (Arcetri) for hospitality and financial support, and to Carlo Baccigalupi and Fabio Finelli for useful discussions and information on the Planck data. Finally, I wish to thank Leonardo Cosmai and Giuseppe Fanizza for their precious help in the preparation of the plots presented in this paper. This work is supported in part by MIUR under grant no. 2012CPPYP7 (PRIN 2012), and by INFN under the program TAsP (Theoretical Astroparticle Physics).

%%%%%%%%%%%%%%%%%%%%%%%%%%%%%%%%%%%%%%
%%%%%%%%%%%%%%%%%%%%%%%%%%%%%%%%%%%%%%

\begin{appendix}

\renewcommand{\theequation}{A.\arabic{equation}}
\setcounter{equation}{0}

\section*{Appendix A. The relevant frequency ratios}
\label{AppA}

Let us compute the frequency ratios needed for an explicit evaluation of the spectral amplitudes of interest for this paper.

We start with $\om_\ast/\om_s$. Using the definition of the string redshift parameter $z_s$ (see Sect. \ref{sec2a}) we can write:
\beq
{\om_\ast\over \om_s}={\om_\ast\over \om_1}{\om_1\over \om_s}= z_s {\om_\ast\over \om_1}.
\label{a1}
\eeq
Noticing that $\om_\ast$ re-enters the horizon in the radiation-dominated era, and taking into account the various phases of the model of post-big bang evolution described in Sect. \ref{sec2b}, we have 
\bea
 {\om_\ast\over \om_1} &\equiv & {H_\ast a_\ast\over H_1 a_1}= 
 {H_\ast\over H_1} \left(a_\ast\over a_d\right)_{\rm rad}
 \left(a_d \over a_\sg\right)_{\rm mat} \left(a_\sg\over a_1\right)_{\rm rad}
 \nonumber \\ &=& 
 {H_\ast\over H_1}\left(H_d\over H_\ast\right)^{1/2}\left(H_\sg\over H_d\right)^{2/3}
\left(H_1\over H_\sg\right)^{1/2}
 \nonumber \\ &=& 
 \left(H_\ast\over \Mp\right)^{1/2}\left(H_1\over  \Mp\right)^{-1/2}\left(m\over \Mp\right)^{-1/3}
\left(\sg_i\over  \Mp\right)^{2/3} \nonumber \\ 
\label{a2}
\eea
(we have used the explicit definitions of $H_d$, $H_\sg$). Finally, we can conveniently refer $\om_\ast$ to the matter-radiation equality scale as $\om_\star \simeq 5 \om_{eq}$, and use \cite{31} $H_{eq} \simeq 0.8 \times 10^{-55} \Mp$, which implies
\beq
 \left(H_\ast\over \Mp\right)^{1/2}\simeq 5  \left(H_{\rm eq}\over \Mp\right)^{1/2} \simeq 10^{-27}.
 \label{a3}
 \eeq
Hence:
\beq
{\om_\ast\over \om_s} \simeq 10^{-27} z_s 
\left(H_1\over  \Mp\right)^{-1/2}\left(m\over \Mp\right)^{-1/3}
\left(\sg_i\over  \Mp\right)^{2/3}.
\label{a4}
\eeq

We follow the same procedure for the other frequency scales appearing in the graviton spectrum (\ref{41}). We start with 
\bea
{\om_d\over \om_s}&=&{\om_d\over \om_1}{\om_1\over \om_s}= z_s {H_d a_d\over H_1a_1}
\nonumber \\
&=& z_s {H_d\over H_1}
 \left(a_d \over a_\sg\right)_{\rm mat} \left(a_\sg\over a_1\right)_{\rm rad},
\label{a5}
\eea
and use the relations $a\sim H^{-2/3}$ for the matter-dominated phase, and $a\sim H^{-1/2}$ for the radiation phase. Inserting the explicit definitions of $H_d$, $H_\sg$ we find 
\beq
{\om_d\over \om_s} \simeq  z_s 
\left(H_1\over  \Mp\right)^{-1/2}\left(m\over \Mp\right)^{7/6}
\left(\sg_i\over  \Mp\right)^{2/3}.
\label{a6}
\eeq
In the same way we also find
\beq
{\om_d\over \om_\sg}= {H_d a_d\over H_\sg a_\sg}=
\left(H_d\over  H_\sg\right)^{1/3}=
\left(m\over \Mp\right)^{2/3}
\left(\sg_i\over  \Mp\right)^{-4/3}.
\label{a7}
\eeq

Let us now consider the constraint imposed by pulsar-timing data, according to Eq. (\ref{44}). We must separately discuss the possibility that the relevant frequency $\om_p \simeq 10^{-8}$ Hz belongs either to the low-energy branch, or to the string branch, or to the axion-dominated branch of the spectrum (\ref{41}). We thus explicitly need three frequency ratios. 

The first is $\om_p/\om_s$, which we can write as
\beq
{\om_p\over \om_s}= {\om_p\over \om_\ast}{\om_\ast\over \om_s},
\label{a8}
\eeq
in order to exploit our previous result (\ref{a4}). We should recall, to this purpose, that the proper frequency $\om_0$ of a mode re-entering today the Hubble horizon is given by \cite{31} $\om_0 \simeq 2.2 \times 10^{-18}$ Hz, and that $\om_\ast \simeq 2.22 \times 10^2 \om_0$. Hence
\beq 
{\om_p\over \om_\ast} \simeq 0.2 \times 10^{8},
\label{a9}
\eeq
so that 
\beq
{\om_p\over \om_s}\simeq 0.2 \times 10^{-19}
z_s 
\left(H_1\over  \Mp\right)^{-1/2}\left(m\over \Mp\right)^{-1/3}
\left(\sg_i\over  \Mp\right)^{2/3}.
\label{a10}
\eeq
What we need, also, is 
\beq
{\om_p\over \om_d}= {\om_p\over \om_s}{\om_s\over \om_d}.
\label{a11}
\eeq
By applying the previous results (\ref{a10}), (\ref{a6}) we obtain
\beq
{\om_p\over \om_d} \simeq 0.2 \times 10^{-19}\left(m\over \Mp\right)^{-3/2}. 
\label{a12}
\eeq
The same procedure also gives
\beq
{\om_p\over \om_\sg}= {\om_p\over \om_d}{\om_d\over \om_\sg} \simeq 0.2 \times 10^{-19}\left(m\over \Mp\right)^{-5/6}
\left(\sg_i\over  \Mp\right)^{-4/3}.
\label{a13}
\eeq

Finally, we need to evaluate the spectral energy density $\Om_{GW}$ at the LIGO sensitivity scale $\om_L \simeq 10^2$ Hz, in order to impose the constraint (\ref{46}).

Following the same procedure as before, and using, as before, 
$\om_\ast \simeq 5 \times 10^{-16}$ Hz, we immediately obtain
\bea
{\om_L\over \om_s}&=& {\om_L\over \om_\ast}{\om_\ast\over \om_s} \simeq 0.2 \times 10^{18}{\om_\ast\over \om_s}
\nonumber \\
&\simeq & 0.2 \times 10^{-9}
z_s 
\left(H_1\over  \Mp\right)^{-1/2}\left(m\over \Mp\right)^{-1/3}
\left(\sg_i\over  \Mp\right)^{2/3}.\nonumber \\ &&
\label{a14}
\eea
Also,
\beq
{\om_L\over \om_d}= {\om_L\over \om_s}{\om_s\over \om_d}
\simeq 0.2 \times 10^{-9}
 \left(m\over \Mp\right)^{-3/2},
\label{a15}
\eeq
and 
\beq
{\om_L\over \om_\sg}= {\om_L\over \om_d}{\om_d\over \om_\sg}=
0.2 \times 10^{-9}
\left(m\over \Mp\right)^{-5/6}
\left(\sg_i\over  \Mp\right)^{-4/3}.
\label{a16}
\eeq

The corresponding frequency ratios for eLISA, i.e. for the frequency scale  $\om_{eL} \simeq 10^{-2}$ Hz, are simply obtained by rescaling Eqs. (\ref{a14})--(\ref{a16}) by the overall factor $10^{-4}$.

\end{appendix}

%%%%%%%%%%%%%%%%%%%%%%%%%%%%%%%%%%%%%%%%%%%%%%%%%%%%%%%%%%%%%%%%%%%%%%%%%%%%%%%%

%%%%%%%%%%%%%%%%%%%%%%%%%%%%%%

\end{document}